\documentclass[fleqn,12pt,twoside]{article}
\usepackage{espcrc1}
\usepackage{epsfig}

\def\al{\alpha}

\def\ga{\gamma} 
\def\de{\delta} \def\De{\Delta}

\def\la{\lambda} \def\La{{\Lambda}}

\def\si{\sigma} \def\Si{{\it\Sigma}}
  
\def\w{\omega}

\def\pa{\partial}

\def\pa{\partial}

\def\ODR{O$\Delta$R}
\def\beq{\begin{equation}}
\def\eeq{\end{equation}}
\def\bea{\begin{eqnarray}}
\def\eea{\end{eqnarray}} 
\def\beqa{\begin{equation}\begin{array}{l}}
\def\eeqa{\end{array}\end{equation}}
\def\eqlab#1{\label{eq:#1}}
\def\figlab#1{\label{fig:#1}}

\def\Eqref#1{Eq.~(\ref{eq:#1})}

\def\Figref#1{Fig.~\ref{fig:#1}}

\def\ol#1{\overline{#1}}

\newcommand{\AmS}{{\protect\the\textfont2
  A\kern-.1667em\lower.5ex\hbox{M}\kern-.125emS}}

\title{Chiral effective field theory of $\Delta$(1232) and Compton
scattering}

\author{Vladimir Pascalutsa and Daniel R.\ Phillips\\[3mm]
Department of Physics and Astronomy,
Ohio University, Athens, OH 45701%
}
      
\begin{document}

\maketitle

\begin{abstract}
\noindent
{\small A new power-counting scheme for a chiral effective field
theory that includes $\De$ degrees of freedom is briefly described
and applied to the nucleon Compton scattering. 
}
\end{abstract}

\bigskip

Chiral perturbation theory ($\chi$PT) is limited to a low-energy domain
$p\ll \La$, where $p$ is the momentum of the particles and
$\La$ stands for all the heavy scales such as the  masses
of the ``integrated out'' degrees of freedom. In a theory without explicit
$\Delta(1232)$ fields the limit
is set by the excitation 
energy of the $\Delta$-resonance: $\De\equiv M_\De-M_N\approx 293$ MeV.
This limit is clearly seen in Compton scattering
off the proton and deuteron, see, e.g., \cite{Bea03}: $\chi$PT  with pions but
without Delta degrees of freedom is in very good agreement
with experiment up to the
photon energy $\w\simeq 100$ MeV, but fails above 150 MeV. 

Recently we have introduced a new power-counting
scheme~\cite{PP} which includes the $\De$ and thus extends 
$\chi$PT into the 
$\De$-resonance region. 
The scheme is set up
such that it is both closely
connected to the usual $\chi$PT without $\De$'s 
(which assumes $m_\pi \ll \De$) 
in the low-energy region $\w\sim m_\pi$ 
{\it and} applies in the $\De$-region
$\w\sim \De$. This is achieved by recognizing the hierarchy of scales:
\beq
\eqlab{scalehi}
m_\pi \ll \De\ll \La \,.
\eeq
This differs dramatically from the older schemes~\cite{JM91,SSE} which
regard $m_\pi$ and $\De$ as similar, $m_\pi\sim\De\ll \La$.
 
With the three-scale hierarchy, \Eqref{scalehi}, one in principle has
two small expansion parameters: 
$m_\pi/\De$ and $\De/\La$. Assuming the breaking-down scale is now
set by the next excitation, $\La\sim 600$ MeV, 
we can regard both of them as roughly the same and introduce a
single small parameter:
\begin{equation}
\delta = \frac{\Delta}{\Lambda} \sim \frac{m_\pi}{\Delta}.
\end{equation}
Each graph can then be characterized 
by an overall $\delta$-counting index, $\al$,
which simply tells us that the graph is of size $\de^\al$. 
The index $\al$ has two different expressions, depending
on whether the photon energy $\omega$ is in the vicinity of $m_\pi$ or
$\De$. For a graph with $L$ loops, $N_\pi$ pion propagators, $N_N$
nucleon propagators, $N_\De$ Delta propagators, and $V_i$ vertices of
dimension $i$, the index is
\beq
\eqlab{alphaHB}
\al = \left\{ \begin{array}{cc} 2 \al_{\chi{\mathrm PT}} - N_\De\,, & \w\sim m_\pi; \\
	 \al_{\chi{\mathrm PT}} - N_\De\,, & \w\sim \De, \end{array}\right.
\eeq
where
$ \alpha_{\chi{\mathrm PT}}=\sum_i i V_i -2  + 4 L  - N_N - 2 N_\pi $
is the index in $\chi$PT with no $\De$'s.

In the region $\w\sim\De$, there is an important exception to this
scaling rule: graphs that are one-Delta-reducible (O$\De$R), such
as those in \Figref{fig1}(c), scale as
$
\delta^\alpha /\left({\omega - \Delta}\right)^{N_{O\De R}},
$
where $N_\De$ in \Eqref{alphaHB} now counts only the
one-Delta-irreducible propagators, while $N_{O\De R}$ is the number of
\ODR\ propagators. At $\w\simeq \De$ the \ODR\ graphs 
all become large and hence need to resummed. 
This amounts to dressing the propagator of the $\De$, thus
ameliorating the divergence which otherwise occurs at the $\De$ pole.
Once dressing
is performed a \ODR\ graph can have only one Delta propagator, and such
a graph then scales as:
\beq
\delta^\alpha \left(\frac{1}{\omega - \Delta -\Si }\right),
\label{eq:ODRscalingnew}
\eeq
where $\Si$ is the self-energy of the $\Delta$. 
The expansion for $\Si$ begins at
$\de^3$, and so the
correct index of an \ODR\ graph in the region $\w\sim\De$ is
\beq
\eqlab{alODR}
\al=\al_{\chi{\rm PT}}-N_\De -2\,,
\eeq 
thus enhanced by two negative powers of $\de$.
As a result, the largest of the \ODR\ graphs
in \Figref{fig1}(c) are
promoted from  next-next-to-leading order (NNLO) 
in the low-energy region to leading order (LO) in the $\De$ region.
\begin{figure}[tb]
\vskip-2mm
\centering
\parbox{15cm}{
\resizebox{15cm}{!}{{\Large (a)}\includegraphics*[0.8cm,21.1cm][19cm,22.6cm]{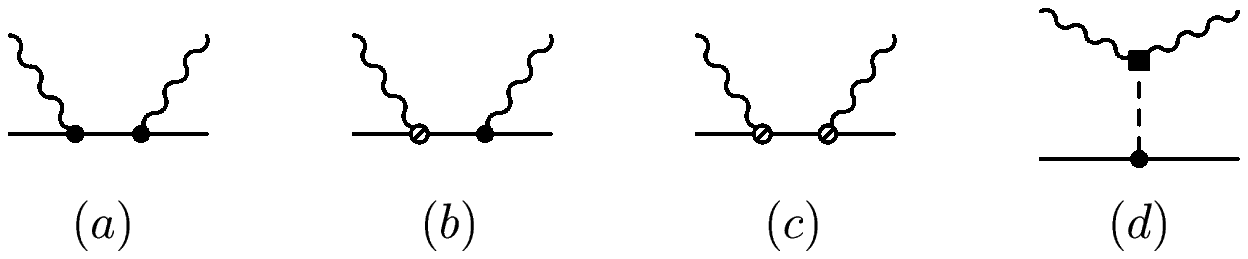}  }}
\parbox{15cm}{
\resizebox{15cm}{!}{{\Large (b)}\includegraphics*[1.9cm,21.5cm][22cm,23cm]{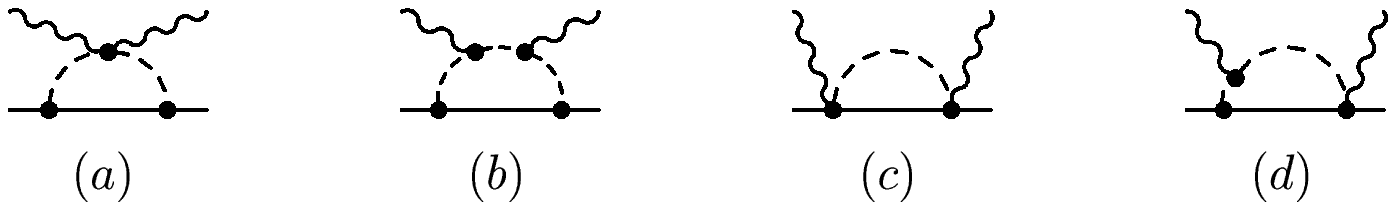}  }}
\parbox{15cm}{
\resizebox{15cm}{!}{{\Large (c)}\includegraphics*[1.1cm,21.5cm][19cm,23cm]{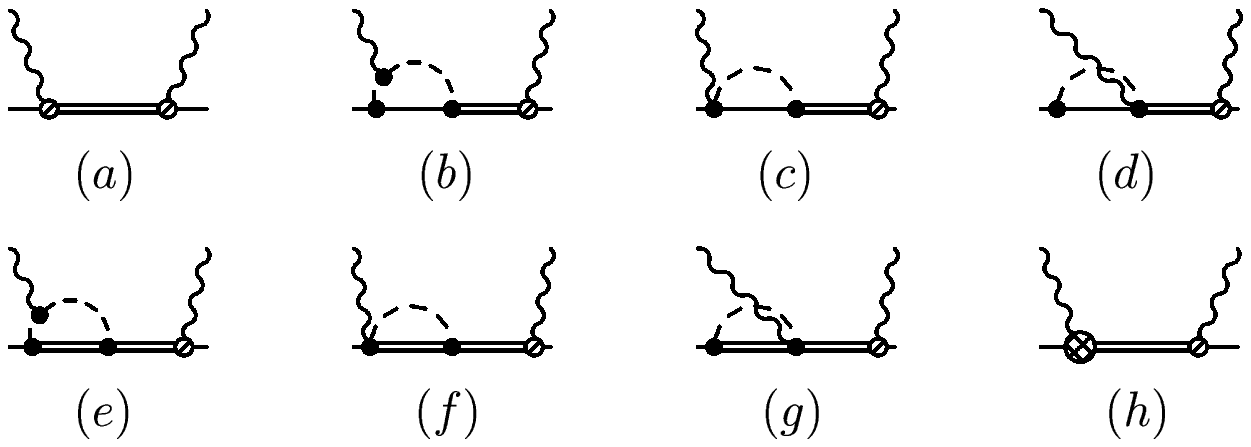} }}
\vskip-5mm
\caption{Graphs computed to obtain a complete NLO 
Compton amplitude in the $\de$-expansion (crossed
partners are not shown, but are 
included).}
\vskip-5mm
\figlab{fig1}
\end{figure}
For the case of Compton scattering on nucleons
the complete next-to-leading order (NLO)
result in both the low-energy and the $\De$ region is
given by the graphs shown in \Figref{fig1}. 
This calculation has three free parameters associated with
the $\pi N\De$ and $\ga N\De$ vertices:
\bea
\eqlab{pinDe}
{\cal L}_{\pi N\De}&=& \frac{i h_A}{2 f_\pi M_\De}\,
\ol N\, T_a^\dagger \,\ga^{\mu\nu\la}\, (\pa_\mu \De_\nu)\, \pa_\la \pi^a 
+ \mbox{H.c.}\, ,\\
\eqlab{ganDe}
{\cal L}_{\ga N \De}&=& \frac{3\,e}{2M_N(M_N+M_\Delta)}\,\ol N\, T_3^\dagger
\left(i g_M  \tilde F^{\mu\nu}
- g_E \gamma_5 F^{\mu\nu}\right)\,\pa_{\mu}\De_\nu
+ \mbox{H.c.}
\eea
where $T_a$ are the isospin 1/2 to 3/2 transition matrices,
see Refs.~\cite{PP,Pas01} for further details.
The relation of the $\ga N\De$ couplings $g_M$ and $g_E$
to the helicity and multipole amplitudes is given in the
Appendix of Ref.~\cite{PP03b}.
 
The $\pi N\De$
coupling $h_A$ is fixed from the $\De$  width ($h_A\simeq 2.8$)
 while the two $\ga N\De$
couplings are adjusted ($g_M=2.6$, $g_E=-6$) to reproduce
the experimental data on proton Compton scattering in the $\Delta$
region. Our NLO result agrees with experiment in a broad energy range.
Examples of this agreement are displayed in \Figref{fig2}. 
\begin{figure}[htb]
\centering
\includegraphics[width=11cm]{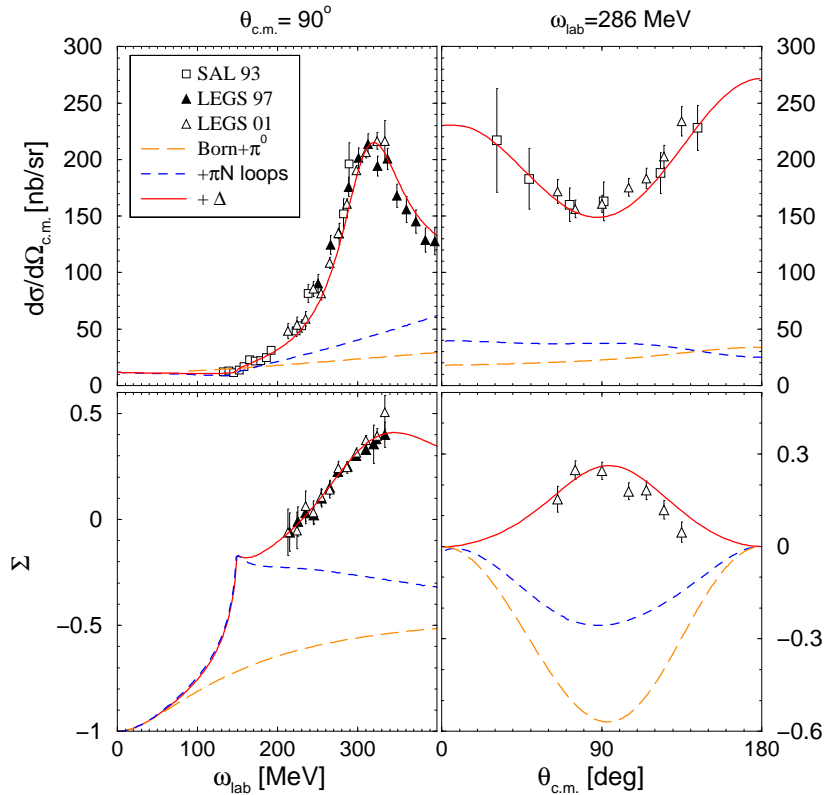}
\vspace{-8mm}
\caption{\small 
The $\de$-expansion next-to-leading order calculation of the $\ga$p
differential cross section ($d\si/d\Omega_{c.m.}$)
and the photon beam asymmetry ($\Sigma$) compared with recent experimental
data: SAL93~\cite{Hal93}, LEGS97~\cite{LEGS97}, LEGS01~\cite{LEGS01}. 
Left panel: energy dependence at a fixed scattering angle. Right panel: angular
dependence at a fixed energy. The long-dashed orange line represents the sum of nucleon and pion
Born graphs, the blue dashed line gives the NLO $\chi$PT prediction, and
the red solid line is the full result at NLO in the $\delta$-expansion.}
\vspace{-5mm}
\figlab{fig2}
\end{figure}

In conclusion, we have presented a new power-counting scheme
which incorporates the $\De$ degrees of freedom into $\chi$PT.
The scheme is based on a three-scale hierarchy of a low- ($m_\pi$),
intermediate- ($\De$), and high- ($\La$) energy scales. In this scheme
the $\De$-resonance width arises naturally as the result of the enhancement
of the \ODR\ graphs in the intermediate-energy regime. The resulting
effective field theory has a significantly broader energy-range of applicability
than $\chi$PT as has been demonstrated here on the example of Compton scattering.

\end{document}